# CREEPING FLOW OF MICROPOLAR FLUID THROUGH A SWARM OF CYLINDRICAL CELLS WITH POROUS LAYER (MEMBRANE)

D. Yu. Khanukaeva*[1], A. N. Filippov[1], P. K. Yadav[2], A. Tiwari[3]

[1]*Gubkin Russian State University of Oil and Gas (National Research University)
Leninsky prospect, 65-1, Moscow, 119991, Russia*
[2] *Motilal Nehru National Institute of Technology Allahabad, Allahabad -211004, India*
[3] *Birla Institute of Technology & Science, Pilani-333031, Rajasthan, India*

*e-mail: khanuk@yandex.ru*

**Abstract:** The flow of micropolar fluid through a membrane modeled as a swarm of solid cylindrical particles with porous layer using the cell model technique is considered. The flow is directed perpendicular to the axis of the cylinders. Boundary value problem involves traditional conditions of velocities and stresses continuity, no-stress and no-couple stress / no-spin condition on hypothetical cell surface. The problem was solved analytically. The influence of micropolar and porous medium parameters on hydrodynamic permeability of a membrane has been studied.

## Introduction

The processes of membrane filtration, flows through sand beds and oil collectors, etc. represent a range of applications of flow problems through random assemblage of particles. Along with Darcy [1] and Brinkman [2] formulations for modeling of flows through porous media, the Happel-Brenner cell model technique [3] is widely used [4-13]. This approach implies taking one particle in the swarm and posing it in a hypothetical cell. The effect of neighboring particles is modeled by applying appropriate boundary condition on the cell surface. So, the problem is reduced to a flow in a single cell, having usually cylindrical or spherical shape. Modern models consider also spheroid cell shapes and different variations of particle materials [14]. Solid, porous and liquid particles confined by liquid envelope are used for flow modeling in different materials. Partially porous particle consisting of a rigid core, covered with a porous non-deformable hydrodynamically uniform layer, in the unbounded incompressible liquid was originally considered in [5]. Various types of boundary conditions are used on solid surfaces and outer hypothetical cell surface Happel [15, 16], Kuwabara [17], Mehta-Morse [18], Kvashnin [19]. Condition of the flow uniformity that is addressed usually as the Mehta-Morse condition was initially proposed by Cunningham in 1910 [20]. Thus far none of them was proved to be preferable than the others.

The overwhelming majority of works on cell models deal with Newtonian fluids. Meanwhile, such class of non-Newtonian fluids as micropolar liquids possess wide opportunities for practical applications and researches due to the existence of analytical solutions both for free and for filtration flows. Nevertheless, only few applications of micropolar theory in cell models were published [21-24] and, up to the knowledge of the authors, combined solid-porous particle in the cell has not been considered anywhere. The mathematical theory of micropolar flows was developed by Eringen [25, 26]. The review of analytical solutions for classical problems in the frame of simple microfluids model and basic applications is given by Khanukaeva and Filippov in [27]. Also, the mentioned review contains formulated boundary value problems for cylindrical and spherical cells with solid core, porous layer and micropolar liquid layer. The problem of flow along the axis of cylindrical cell is solved and analyzed in our previous work [28]. This paper continues the study of micropolar flow through the assemblage of cylindrical particles by the consideration of flow perpendicular to the cell axis. Chaotic orientation of cylinders is also discussed.

After solving the problem, the hydrodynamic permeability of a membrane regarded as a system of impermeable particles covered with a porous layer is calculated. The review of the hydrodynamic permeability study for different cell constructions and flow regimes is given in [28]. This integral characteristic of the flow is used for the study of boundary conditions influence on the process and for all necessary parametric investigations.

## 1. Statement of the problem

The uniform flow of velocity **U** is directed perpendicular to the symmetry axis of the cell, which consists of three coaxial layers as it is shown in Fig.1. The inner layer is a solid core of radius $a$, the intermediate layer $a < r < b$ is porous, and the outer layer $b < r < c$ is occupied by the free micropolar liquid. The cylindrical coordinate system $(r, \theta, z)$ is introduced so that the direction of vector **U** corresponds to $\theta = 0$.

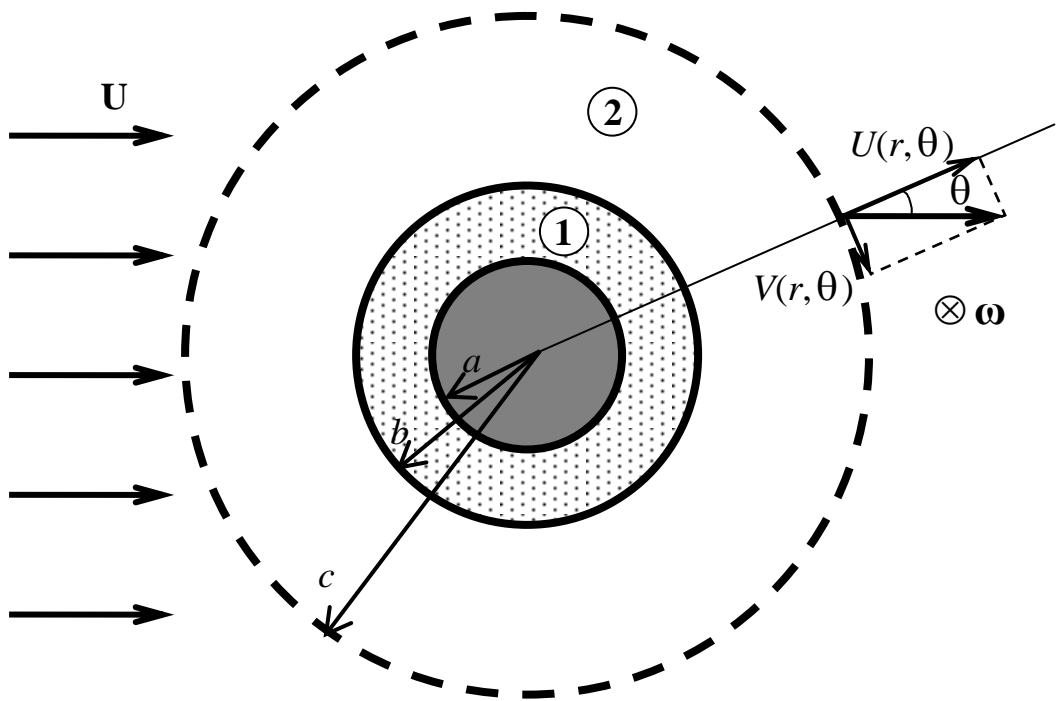

Figure 1. The scheme of the flow.

The theory of micropolar liquids, which describes the flow in region 2, was developed by Eringen [26, 29]; field equations of creeping flow include the continuity equation, the momentum equation and the moment of momentum equation:

$$\nabla \cdot \mathbf{v} = 0,$$
$$\mathbf{0} = \rho \mathbf{F} - \nabla P + (\mu + \kappa)\Delta \mathbf{v} + 2\kappa \nabla \times \boldsymbol{\omega},$$
$$\mathbf{0} = \rho \mathbf{L} + (\alpha + \delta - \varsigma)\nabla \nabla \cdot \boldsymbol{\omega} + (\delta + \varsigma)\Delta \boldsymbol{\omega} + 2\kappa \nabla \times \mathbf{v} - 4\kappa \boldsymbol{\omega},$$

where $\mathbf{v}, \boldsymbol{\omega}$ are linear and angular velocity vectors correspondingly, $P$ is the pressure, $\rho$ is the liquid density, $\mathbf{F}, \mathbf{L}$ are the densities of external forces and couples, $\mu, \kappa, \alpha, \delta, \varsigma$ are the viscosity coefficients of the micropolar medium. The notation of viscosity coefficients used here slightly differs from the original notation of Eringen. It is chosen in the way that coefficient $\mu$ is equal to a dynamic viscosity of the Newtonian liquid. Rotational viscosity $\kappa$ relates skew symmetrical part of the deformation rate tensor $\hat{\gamma}$ with the stress tensor $\hat{t}$. Angular viscosities $\alpha, \delta, \varsigma$ are the coefficients in the constitutive equation relating the curvature-twist rate tensor $\hat{\chi}$ with the couple stress tensor $\hat{m}$. In the linear theory of simple micro fluids, the stress tensor and the couple stress tensor can be expressed via the deformation rate tensor and the curvature-twist rate tensor in the form accepted in the micropolar theory of elasticity [30]

$$\hat{t} = (-P + \lambda \operatorname{tr}\hat{\gamma})\hat{G} + 2\mu\hat{\gamma}^{(S)} + 2\kappa\hat{\gamma}^{(A)},$$
$$\hat{m} = \alpha(\operatorname{tr}\hat{\chi})\hat{G} + 2\delta\hat{\chi}^{(S)} + 2\varsigma\hat{\chi}^{(A)},$$

where $\hat{G}$ is the metric tensor, superscripts $(S)$ and $(A)$ denote symmetrical and skew symmetrical parts of tensors correspondingly. The deformation rate tensor and the curvature-twist rate tensor are represented via the gradients of linear and angular velocities correspondingly $\hat{\gamma} = (\nabla \mathbf{v})^T - \hat{\varepsilon} \cdot \boldsymbol{\omega}$, $\hat{\chi} = (\nabla \boldsymbol{\omega})^T$, $\hat{\varepsilon}$ is the Levi-Civita tensor. As one can see, coefficient $\lambda$ may be omitted for incompressible fluids due to the construction of the deformation rate tensor.

The stationary filtration flows are sufficiently slow for the Stokes approach to be valid. In absence of external forces and couples the governing equations for free micropolar fluid can be written as ($b < r < c$)

$$\begin{aligned} &\nabla \cdot \mathbf{v}_2 = 0, \\ &-(\mu + \kappa)\nabla \times \nabla \times \mathbf{v}_2 + 2\kappa \nabla \times \boldsymbol{\omega}_2 = \nabla P_2, \\ &-(\delta + \varsigma)\nabla \times \nabla \times \boldsymbol{\omega}_2 + 2\kappa \nabla \times \mathbf{v}_2 - 4\kappa \boldsymbol{\omega}_2 = 0, \end{aligned} \quad (1)$$

where it was used the equality $\nabla \times \nabla \times \mathbf{a} = \nabla \nabla \cdot \mathbf{a} - \Delta \mathbf{a}$, continuity equation and symmetry of the geometry, which gives $\nabla \cdot \boldsymbol{\omega} = 0$. Subscript 2 is ascribed for all variables in the free stream layer. Subscript 1 will be used for a porous layer.

The governing equations for a porous region involve values of velocities averaged over the elementary representative volume $V$ and defined as $<\mathbf{A}> = \frac{1}{V} \int_V \mathbf{A}\, dV$. While it is recommended [31] to use the pressure averaged over the volume of pores, i.e. the volume occupied by fluid $V_f$, $<P>_f = \frac{1}{V_f} \int_{V_f} P\, dV$, the averaged pressure in the equation of motion can be expressed as $<P> = \varepsilon <P>_f$, where $\varepsilon = V_f / V$ is the porosity of the medium. The standard averaging technique, applied by Kamel et al. [32], led to the equations of Brinkman-type governing the stationary filtration of micropolar liquid:

$$\nabla \cdot \mathbf{v} = 0,$$

$$\nabla P = \left(\frac{\mu}{\varepsilon} + \frac{\kappa}{\varepsilon}\right) \Delta \mathbf{v} + \frac{2\kappa}{\varepsilon} \nabla \times \boldsymbol{\omega} - \frac{\mu + \kappa}{k} \mathbf{v},$$

$$0 = (\alpha + \delta - \varsigma)\nabla <\nabla \cdot \boldsymbol{\omega}> + (\delta + \varsigma)\Delta \boldsymbol{\omega} + 2\kappa \nabla \times \mathbf{v} - 4\kappa \boldsymbol{\omega},$$

where $k$ is the permeability of the porous medium. Making use of the divergence free property of the spin field for the considered geometry of the flow and the abovementioned vector equality one can write the governing equations for porous region as ($a < r < b$)

$$\nabla \cdot \mathbf{v}_1 = 0,$$

$$-\left(\frac{\mu}{\varepsilon} + \frac{\kappa}{\varepsilon}\right) \nabla \times \nabla \times \mathbf{v}_1 + \frac{2\kappa}{\varepsilon} \nabla \times \boldsymbol{\omega}_1 - \frac{\mu + \kappa}{k} \mathbf{v}_1 = \nabla P_1, \qquad (2)$$

$$-(\delta + \varsigma)\nabla \times \nabla \times \boldsymbol{\omega}_1 + 2\kappa \nabla \times \mathbf{v}_1 - 4\kappa \boldsymbol{\omega}_1 = 0.$$

If the non-dimensional variables and values are introduced as follows

$$\tilde{r} = \frac{r}{b}, \quad \ell = \frac{a}{b}, \quad m = \frac{c}{b}, \quad \tilde{\mathbf{v}} = \frac{\mathbf{v}}{U}, \quad \tilde{\boldsymbol{\omega}} = \frac{\omega b}{U}, \quad \tilde{P} = \frac{Pb}{\mu U}, \qquad (3)$$

the non-dimensional forms of systems (1) and (2) are respectively

$$\tilde{\nabla} \cdot \tilde{\mathbf{v}}_2 = 0,$$
$$-(\mu + \kappa)\tilde{\nabla} \times \tilde{\nabla} \times \tilde{\mathbf{v}}_2 + 2\kappa \tilde{\nabla} \times \tilde{\boldsymbol{\omega}}_2 = \mu \tilde{\nabla} \tilde{P}_2, \qquad (4)$$
$$-\frac{\delta + \varsigma}{b^2} \tilde{\nabla} \times \tilde{\nabla} \times \tilde{\boldsymbol{\omega}}_2 + 2\kappa \tilde{\nabla} \times \tilde{\mathbf{v}}_2 - 4\kappa \tilde{\boldsymbol{\omega}}_2 = 0,$$

and

$$\tilde{\nabla} \cdot \tilde{\mathbf{v}}_1 = 0,$$
$$-\frac{\mu + \kappa}{\varepsilon} \tilde{\nabla} \times \tilde{\nabla} \times \tilde{\mathbf{v}}_1 + \frac{2\kappa}{\varepsilon} \tilde{\nabla} \times \tilde{\boldsymbol{\omega}}_1 - \frac{\mu + \kappa}{k} b^2 \tilde{\mathbf{v}}_1 = \mu \tilde{\nabla} \tilde{P}_1, \qquad (5)$$
$$-\frac{\delta + \varsigma}{b^2} \tilde{\nabla} \times \tilde{\nabla} \times \tilde{\boldsymbol{\omega}}_1 + 2\kappa \tilde{\nabla} \times \tilde{\mathbf{v}}_1 - 4\kappa \tilde{\boldsymbol{\omega}}_1 = 0.$$

It is worth mentioning that viscosities $\mu$ and $\kappa$ have one and the same dimension, so number of micropolarity, $N^2 = \kappa/(\mu + \kappa)$, introduced in [33] is non-dimensional. The ratio $(\delta + \varsigma)/\mu$ has the dimension of length squared, so the combination of values $L^2 = \dfrac{\delta + \varsigma}{4\mu b^2}$ [33] represents the relation between the micro and macro scales of the problem. With these two non-dimensional parameters being used, systems (4) and (5) can be rewritten as (tildes are omitted here and further)

$$\nabla \cdot \mathbf{v}_2 = 0,$$
$$-\frac{1}{N^2} \nabla \times \nabla \times \mathbf{v}_2 + 2\nabla \times \boldsymbol{\omega}_2 = \left(\frac{1}{N^2} - 1\right)\nabla P_2, \qquad (6)$$
$$-L^2 \nabla \times \nabla \times \boldsymbol{\omega}_2 + \frac{1}{2}\frac{N^2}{1-N^2} \nabla \times \mathbf{v}_2 - \frac{N^2}{1-N^2} \boldsymbol{\omega}_2 = 0,$$

and

$$\nabla \cdot \mathbf{v}_1 = 0,$$

$$-\frac{1}{N^2}\nabla \times \nabla \times \mathbf{v}_1 + 2\nabla \times \mathbf{\omega}_1 - \frac{\varepsilon\sigma^2}{N^2}\mathbf{v}_1 = \varepsilon\left(\frac{1}{N^2}-1\right)\nabla P_1, \qquad (7)$$

$$-L^2\nabla \times \nabla \times \mathbf{\omega}_1 + \frac{1}{2}\frac{N^2}{1-N^2}\nabla \times \mathbf{v}_1 - \frac{N^2}{1-N^2}\mathbf{\omega}_1 = 0,$$

where the ratio of macro scale of the cell $b$ to the micro scale of porous medium $\sqrt{k}$ is denoted as $\sigma = b/\sqrt{k}$.

General solutions of systems (6) and (7) can be obtained separately and presented in the form $\mathbf{v}_i(r,\theta) = \{u_i(r)\cos\theta; v_i(r)\sin\theta; 0\}$, $\mathbf{\omega}_i(r,\theta) = \{0; 0; \omega_i(r)\sin\theta\}$, $P_i(r,\theta) = p_i(r)\cos\theta$, $i=1,2$ due to the symmetry of the flow. So, each of the systems (6-7) reduces to four scalar equations with respect to four unknown functions of one independent variable: three of these equations are of the second order and one equation is of the first order. Thus, each general solution will contain six arbitrary constants. It means, twelve conditions are required for the closure of the boundary value problem.

The no-slip and no-spin conditions on solid surfaces were essentially used in the derivation of filtration equations (2) [32]. It means, that there is no choice of the conditions on the boundary $r = \ell$. Only no-slip and no-spin are allowed there

$$u_1(\ell) = 0, \quad v_1(\ell) = 0, \quad \omega_1(\ell) = 0. \qquad (8)$$

The liquid-porous interface offers wider variety of conditions. From mechanical point of view the most natural conditions are the continuity of all velocity's components, i.e.

$$u_1(1-0) = u_2(1+0), \quad v_1(1-0) = v_2(1+0), \quad \omega_1(1-0) = \omega_2(1+0) \qquad (9)$$

and continuity of the stress and couple stress tensor components, normal and tangential to the boundary surface. The corresponding components of the stress and couple stress tensors in the chosen coordinate system are

$$t_{rr} = (-p(r) + 2\mu u'(r))\cos\theta,$$

$$t_{r\theta} = \left((\mu+\kappa)v'(r) - (\mu-\kappa)\frac{u(r)+v(r)}{r} - 2\kappa\omega(r)\right)\sin\theta,$$

$$m_{rz} = (\delta+\varsigma)\omega'(r)\sin\theta.$$

The averaging procedure used for the derivation of filtration equations demonstrates that the viscous terms have the coefficients equal to the viscosities of pure liquid divided by the porosity - the so-called effective viscosities. They should be used in the expressions for stress and couple stress components in the porous region. In addition, applying relations (3) one obtains the non-dimensional forms of boundary conditions for stresses and couple stresses:

$$-p_1(1-0) + \frac{2}{\varepsilon}u_1'(1-0) = -p_2(1+0) + 2u_2'(1+0), \tag{10}$$

$$\frac{1}{\varepsilon}v_1'(1-0) - \frac{1-2N^2}{\varepsilon}\frac{u_1(1-0)+v_1(1-0)}{1-0} - 2\frac{N^2}{\varepsilon}\omega_1(1-0) =$$
$$= v_2'(1+0) - (1-2N^2)\frac{u_2(1+0)+v_2(1+0)}{1+0} - 2N^2\omega_2(1+0), \tag{11}$$

$$\omega_1'(1-0) = \varepsilon\omega_2'(1+0). \tag{12}$$

Three more boundary conditions are required at $r = m$. One of them is the continuity of a normal component of the linear velocity:

$$u_2(m) = 1. \tag{13}$$

Four types of conditions are known in classical cell models for non-polar liquids as the second condition at the outer boundary of the cell. They are Happel's no-stress condition [15], Kuwabara's vorticity free condition [17], the symmetry of velocity profile by Kvashnin [19], and the condition of the flow uniformness by Cunningham [20] (frequently called as the Mehta and Morse condition [18]). Each of them can be considered for micropolar liquid. Here we use Happel's no-stress condition:

$$v'_2(m-0) - (1-2N^2)\frac{u_2(m-0) + v_2(m-0)}{m-0} - 2N^2\omega_2(m-0) = 0, \quad (14)$$

supplemented with to different conditions in order to study the influence of one boundary condition variation on the solutions. These two alternative conditions are no-couple stress condition

$$\omega'_2(m-0) = 0 \quad (15)$$

and no-spin condition

$$\omega_2(m-0) = 0. \quad (16)$$

These types of conditions were used in our paper on the parallel flow in cylindrical cell [28]. And they are used here in order to compare the results obtained for the perpendicular flow with the results for parallel flow. Then both types of flow will be combined to give an approach to the flow in the medium modeled with the chaotically oriented cylindrical cells.

Other types of boundary conditions on each of the considered boundaries also can be applied. The discussion of them is given in [27].

What is worth noting, that for perpendicular flow the stress and couple stress tensor components responsible for the considered boundary conditions do not contain gyro-viscosities. So, these viscosities could not be collected in two parameters $N$ and $L$, as it took place for parallel orientation of the cell to the flow direction. Therefore, there is no need to introduce any other parameter of micropolar medium, unlike the case of parallel flow [28].

## 2. General solution of the problem

The general solution of system (6) has the form

$$u_2(r) = \frac{C_1}{r^2} + C_2 \ln r + C_3 r^2 + C_4 + \frac{C_5}{r} I_1\left(\frac{N}{L}r\right) + C_6 \text{Meij}\left(\frac{N}{2L}r\right),$$

$$v_2(r) = -r u_2'(r) - u_2(r),$$

$$\omega_2(r) = -\frac{C_2}{r} - 4C_3 r - \frac{C_5}{2L^2} I_1\left(\frac{N}{L}r\right) + \frac{2C_6}{NL} K_1\left(\frac{N}{L}r\right),$$

$p_2(r)$ is found from the $\theta$-projection of the equation of motion in system (6).

The general solution of system (7) is

$$u_1(r) = \frac{C_7}{r^2} + C_8 + C_9 \frac{I_1(\alpha_1 r)}{r} + C_{10} \text{Meij}\left(\frac{\alpha_1}{2}r\right) + C_{11} \frac{I_1(\alpha_2 r)}{r} + C_{12} \text{Meij}\left(\frac{\alpha_2}{2}r\right),$$

$$v_1(r) = -r u_1'(r) - u_1(r),$$

$$\omega_1(r) = \frac{1}{2N^2}\left(1 + \alpha_2^2 L^2\left(1 - \frac{1}{N^2}\right)\right)\left(-\alpha_1^2 C_9 I_1(\alpha_1 r) + 4\alpha_1 C_{10} K_1(\alpha_1 r)\right) +$$

$$+ \frac{1}{2N^2}\left(1 + \alpha_1^2 L^2\left(1 - \frac{1}{N^2}\right)\right)\left(-\alpha_2^2 C_{11} I_1(\alpha_2 r) + 4\alpha_2 C_{12} K_1(\alpha_2 r)\right),$$

$p_1(r)$ is found from the $\theta$-projection of the equation of motion in system (7).

The following notations are used: $I_1(\xi)$, $K_1(\xi)$ are the modified Bessel and Macdonald functions of the first order correspondingly; $\text{Meij}(\xi)$ is the Meijer G-function $\mathrm{Meij}(\xi) = \dfrac{1}{4\pi i}\int \dfrac{\Gamma(1-s/2)\Gamma^2(s/2)}{\Gamma(2-s/2)}\xi^{-s}ds$, $\Gamma(x)$ is the gamma function. Constants $\alpha_1, \alpha_2$ are defined by the system
$$\begin{cases} \alpha_1^2 + \alpha_2^2 = \dfrac{N^2}{L^2} + \varepsilon\sigma^2, \\ \alpha_1^2 \alpha_2^2 = \dfrac{\varepsilon\sigma^2 N^2}{(1-N^2)L^2}. \end{cases}$$

## 3. Solution of the boundary value problems

The graphical representation of the obtained solutions with conditions (8-14) and (15) or (16) is given in Fig.2, where the streamlines (a) and the corresponding profiles of angular velocities (b) are shown. The stream function $\Psi$ is introduced in a usual way

$$u = \dfrac{1}{r}\dfrac{\partial \Psi}{\partial \theta}, \quad v = -\dfrac{\partial \Psi}{\partial r}.$$

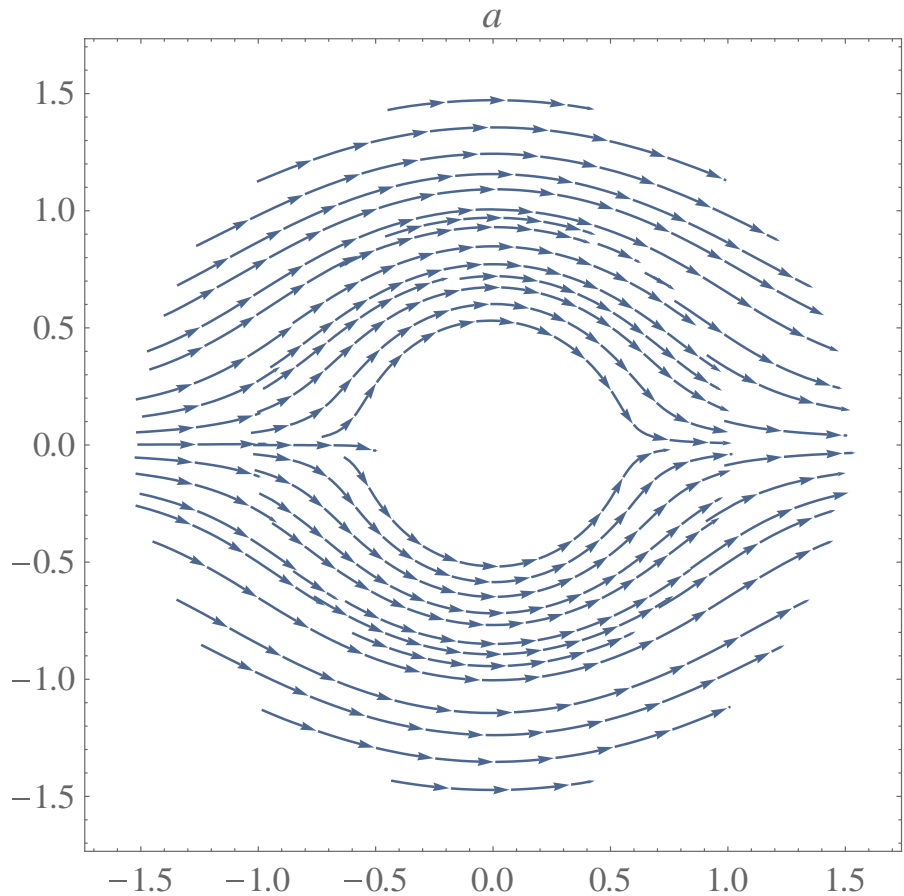

a

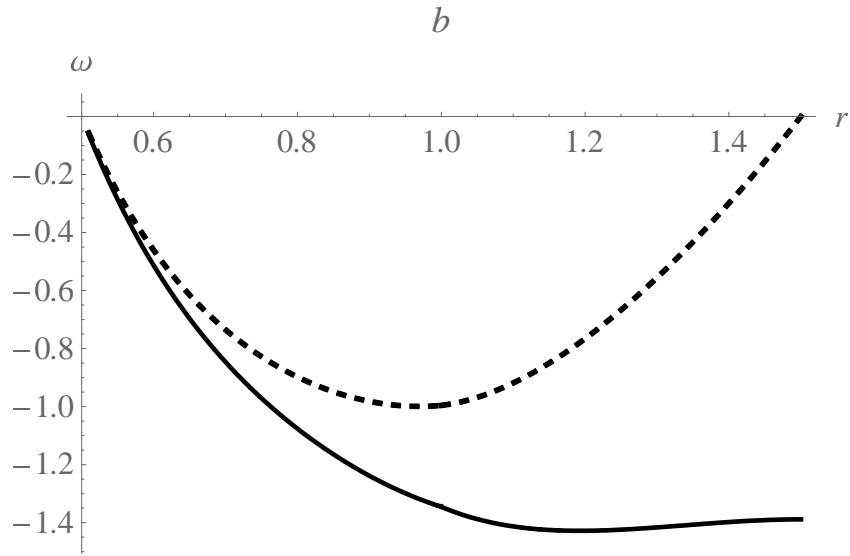

**Fig.2.** The flow field for the no-couple stress condition (a) and variation of angular velocity (b) under the no-couple stress condition (solid lines) and the no-spin condition (dashed lines) on the outer surface of the cell.

Fig.2a shows the flow field for the no-couple stress condition on the outer surface of the cell. It is plotted in non-dimensional units. The annulus shape of the flow domain has inner radius $\ell = 0.5$ and outer radius $m = 1.5$, which represent noticeable thickness of porous layer and middle position ($r = 1$) of porous-liquid interface inside the cell. Characteristic values of parameters used in [28] are taken for plotting Fig.2: $N = 0.5$ and $L = 0.2$ correspond to well-developed micropolarity, $\varepsilon = 0.75$, $\sigma = 3$ characterize porous medium with properties far from any limit cases. The listed values of parameters will be used in the parametric studies below. The streamlines for the no-spin condition are not distinguishable for an unaided eye from those, presented in Fig.2a, therefore they are now demonstrated here. Two curves are given in Fig.2b and in each of the following figures below. A solid line corresponds to the solution with no-couple stress condition at the outer surface of the cell (15), dashed line corresponds to the no-spin condition (16). Due to the definition of the non-dimensional velocity, the values of the linear velocity components do not exceed unity, and one can notice from Fig.2b that the order of the average absolute value of $\omega$ is the same as the order of the linear velocity average absolute value. But for parallel flow these magnitudes differ approximately by two times, the relation between non-dimensional values of ω and *u* being the same. Assumingly this effect is due to the curvature of streamlines, which are for parallel flow are straight-linear and for perpendicular flow are curved and shown in Fig.2a contributing to the rotational effects.

## 4. Results and discussion

In order to analyze the influence of the model parameters on the overall flow, we calculate coefficient of hydrodynamic permeability $L_{11}$. It is defined as $L_{11} = \dfrac{U}{F/V}$, where denominator represents the cell pressure gradient, $F$ is the force which acts on the particle from the flow side, $V = \pi c^2$ is the volume of the cell. Both magnitudes are normalized by the unit of a length. Force $F$ in turn is calculated by the integration of stresses over the outer surface of porous layer

$$F = \oiint_S (t_{rr} \cos\theta - t_{r\theta} \sin\theta) ds.$$

Non-dimensional form of hydrodynamic permeability is then,

$$L_{11} = \dfrac{\pi m^2}{\int_0^{2\pi}\left([-p_2 + 2u_2']\cos^2\theta - \dfrac{1}{1-N^2}\left[v_2' - (1-2N^2)\dfrac{u_2+v_2}{r} - 2N^2\omega_2\right]\sin^2\theta\right)d\theta} = \dfrac{m^2}{4C_2}.$$

The dependencies of the hydrodynamic permeability on the governing parameters are studied below. Each figure contains the curves for two flows: perpendicular to the cell axis, obtained in the present study, and along the cell axis, which was studied in [28].

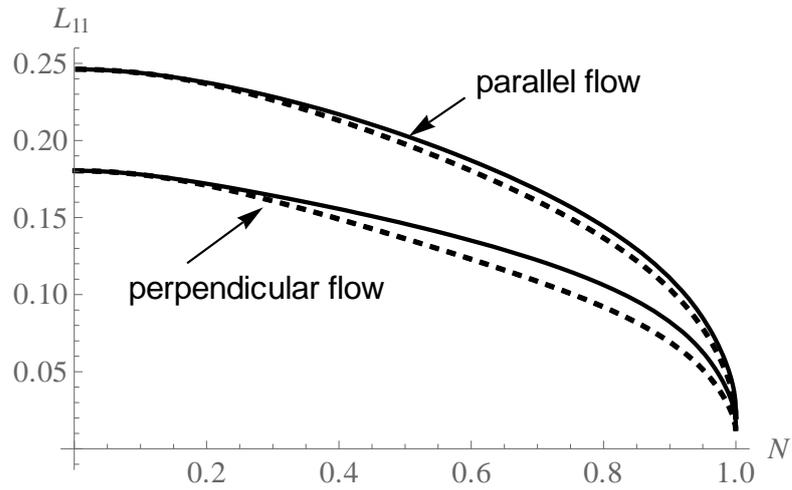

**Fig.3.** Variation of hydrodynamic permeability with coupling parameter $N$ under no-couple stress condition (solid line) and no-spin condition (dashed line) on the outer surface of the cell for parallel and perpendicular flows.

From Fig.3 it is seen, that in all cases the hydrodynamic permeability significantly decays with $N$. The relative variation of $L_{11}$ has the order of 400%, for parallel flow being even higher. The absolute values of $L_{11}$ for parallel flow is also higher than for perpendicular flow. This effect is known for analogous flows of Newtonian liquids too [10] and can be qualitatively explained by the geometrical consideration of parallel package of cells along and perpendicular to the flow. The area of filtration is proportional to the squared radius of the cell in the first case, and to the linear size of the cell in the second case. Therefore, for parallel flow, the velocity and hydrodynamic permeability are higher. Another observation is that all formulations give almost the same behavior of $L_{11}$ for limiting cases of $N$. When micro-level effects are negligible ($N \to 0$) hydrodynamic permeability tends to the corresponding limits for Newtonian fluid and goes almost to zero when these effects are extremely exhibited ($N \to 1$). Limiting case of $N \to 1$ is never reached, because microrotational viscosity cannot tend to infinity. The last note should be made on the effect of different boundary conditions upon the hydrodynamic permeability. Again, the curves corresponding to the no-spin boundary condition are located lower, like in case of parallel flow, because the velocity of the flow is lower in this case, but the discrepancy of the curves calculated for condition (15) and condition (16) is not large.

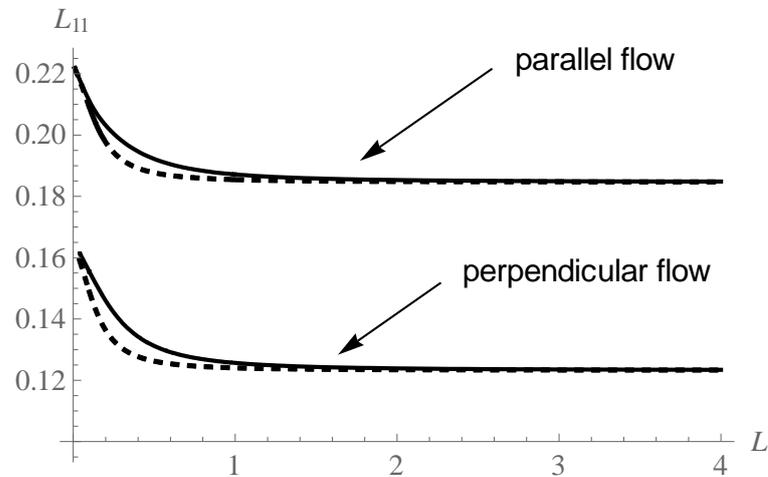

**Fig.4.** Variation of hydrodynamic permeability with scale parameter $L$ under no-couple stress condition (solid line) and no-spin condition (dashed line) on the outer surface of the cell for parallel and perpendicular flows.

Fig.4 shows that the hydrodynamic permeability of the membrane decreases with the increase of the scale parameter $L$: the higher the microscale of the liquid the more it suppresses the flow. Analogous to the dependence of $L_{11}$ on $N$, one can see in Fig.4 higher absolute values of hydrodynamic permeability for a membrane composed of cylinders with axes parallel to the flow direction. But for perpendicular orientation of cylinders the relative variation of $L_{11}$ is more substantial and reaches 100%. The asymptotic behavior of the hydrodynamic permeability as a function of $L$, mentioned in [28], is governed by the value of $N$. As it follows from Fig.4, the cell orientation does not influence the position of asymptote and determines only its absolute value. Again, in both cases a very small variation of the hydrodynamic permeability is obtained depending on the type of boundary condition at the outer surface of the cell.

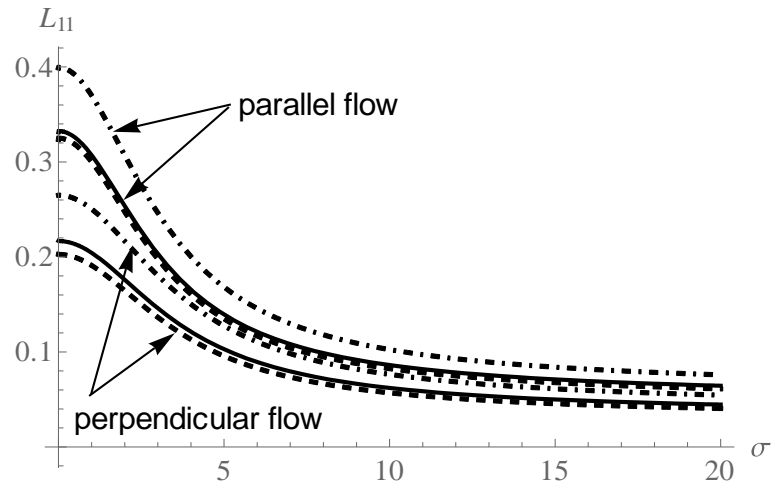

**Fig.5.** Variation of hydrodynamic permeability with parameter σ under no-couple stress condition (solid line), no-spin condition (dashed line) on the outer surface of the cell and for the Newtonian liquid (dot-dashed line) for parallel and perpendicular flows.

The effect of permeability parameter σ on the hydrodynamic permeability of membrane is shown in Fig. 5. For all values of σ the hydrodynamic permeability for the perpendicular orientation of cylinder is lower, than for parallel flow. A rise in dimensionless permeability parameter σ leads to a reduction of the membrane permeability for all models, boundary conditions and geometries of the flow, as well as it was observed for a Newtonian liquid [10]. This behavior is quite natural due to the definition of σ. It is reversely proportional to the square root of the porous medium permeability, called Brinkman radius. It represents the characteristic scale of filtration flow. So, σ~1 corresponds to the flow in the whole volume of the cell, for one order lower values of σ the presence of porous layer is almost negligible, which results in a significant hydrodynamic permeability of the membrane. On the other side, high values of σ imply so small permeability coefficient of the porous medium $k$, that the filtration occurs in a thin layer, that does not influence the overall flow and the hydrodynamic permeability $L_{11}$ is determined only by a free flow layer $1 < r < m$. At last, Fig.5 once more demonstrates, that the value of hydrodynamic permeability of the membrane is slightly higher for no-couple stress condition that for no-spin condition.

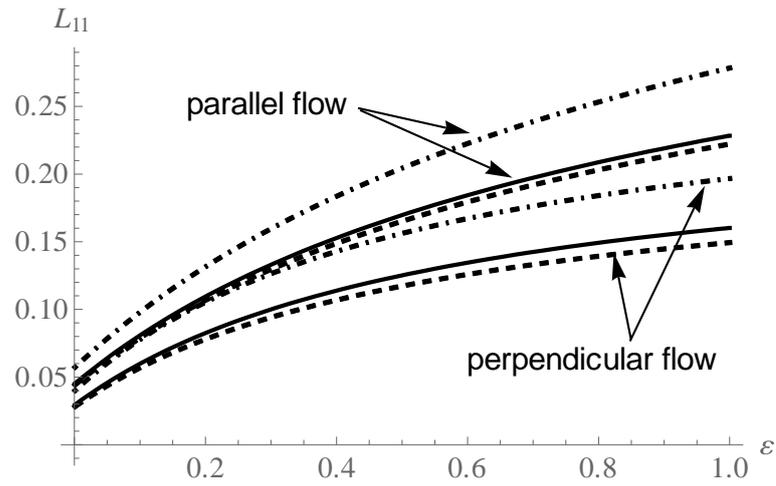

**Fig.6.** Variation of hydrodynamic permeability with porosity ε under no-couple stress condition (solid line), no-spin condition (dashed line) on the outer surface of the cell and for the Newtonian liquid (dot-dashed line) for parallel and perpendicular flows.

The behavior of the hydrodynamic permeability of a membrane with increasing porosity ε of the layer $\ell < r < 1$ is shown in Fig.6. It demonstrates quite natural growth of the hydrodynamic permeability $L_{11}$ with the increase of porosity. One can notice higher discrepancy of the hydrodynamic permeability values for Newtonian and micropolar flows at large porosity magnitudes. Although, strictly speaking, this dependence should not be analyzed at low values of ε, when the validity of Brinkman approach is not proved.

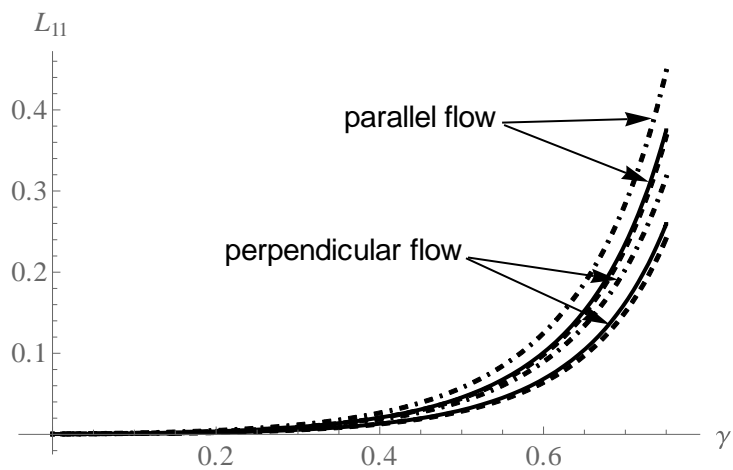

**Fig.7.** Variation of hydrodynamic permeability with porosity γ under no couple stress condition (solid line), no spin condition (dashed line) on the outer surface of the cell for parallel and perpendicular flows.

The discussed porosity ε is the characteristics of the inner porous layer of the cell, while it is known the overall porosity of the membrane as a whole γ. It is usually defined as the ratio of volume of voids to the whole volume of the membrane. For the structure of the membrane considered in the present paper, it can be calculated as the sum of free space of the cell and the volume of pores in the layer $\ell < r < 1$:

$$\gamma = \frac{m^2 - 1 + \varepsilon(1 - \ell^2)}{m^2} = 1 - \frac{1}{m^2} + f(\varepsilon, \ell, m).$$ Thus, in general case it depends on the cell layers sizes and intrinsic porosity ε. The thinner the porous layer of the cell (the closer the size $\ell$ to unity) the less the term $f(\varepsilon, \ell, m)$. So, for high values of $\ell$ $\gamma \approx 1 - 1/m^2$. The dependence of the hydrodynamic permeability on this parameter is shown in Fig.7, where $\ell = 0.9$ was used along with the rest parameters stated above. A rapid increase of the hydrodynamic permeability of membrane with rising γ is observed for all models, boundary conditions, orientations of cells and liquid types. This points to the major responsibility of this parameter for the modeling of flows in membranes and secondary role of the type of a liquid model (Newtonian or Non-Newtonian), cells orientation with respect to the flow direction (parallel of perpendicular) and boundary conditions on the outer surface of the cell (no-couple stress or no-spin).

## 5. Conclusion

The present work continues the application of the cell model technique in modeling of filtration flows for the non-Newtonian fluids. The solutions of the boundary value problems were obtained in totally analytical form, which allows parametric study of the flow peculiarities. In the present study the hydrodynamic permeability was taken as a factor of interest representing the integral characteristics of the system. Due to the cumbersome description of the hydrodynamic permeability in the problem under consideration, its dependence upon any parameter cannot be studied analytically. So, a set of computational experiments was fulfilled.

The present study confirms the conclusion made in [28] about almost negligible influence of the considered conditions at the outer surface of the cell, when such integral characteristics as hydrodynamic permeability is studied. The total membrane porosity γ demonstrates the strongest effect on the hydrodynamic permeability, while the intrinsic porosity ε and permeability parameter σ show moderate influence on the hydrodynamic permeability. Nearly the same effect is observed for the variation of micropolar liquid properties in their allowed ranges. All the obtained dependencies of the hydrodynamic permeability for the perpendicular flow are compared with analogous ones for the parallel flow. It is found that the hydrodynamic permeability for the perpendicular flow is lower than for the parallel flow under same flow conditions.

Real fibrous membranes frequently do not represent a package of parallel cylindrical fibers. Most likely, they are oriented chaotically. Happel and Brenner [3] suggested the averaging procedure for the calculation of the membrane hydrodynamic permeability, including the weighted values of $L_{11}$ for parallel and perpendicular orientations of cylinders with respect to the flow direction. The common hydrodynamic permeability is regarded as the sum of permeabilities for parallel and perpendicular flows with coefficients 1/3 and 2/3 respectively. So, the separate solutions of the boundary value problems for parallel and perpendicular flows allow to obtain the hydrodynamic permeability of a real fibrous membrane. In the present paper the curves for chaotic package of cylinders are not shown in the graphs. Instead both curves for parallel and perpendicular orientations of the cells are given, so that the position of the curves for chaotic orientation can be easily predicted between them.

**Acknowledgement:**

The present work is supported by DST (INT/RUS/ RFBR/ P-212) for Indian side and RFBR (19-08-00058) for Russian side.